\documentclass[twocolumn,aps,prc,preprintnumbers,showpacs,showkeys,nofootinbib,superscriptaddress,fleqn,floatfix,tightenlines, 10pt]{revtex4-1}
\usepackage{amsmath,amsfonts,amssymb,amscd,amsxtra,amsthm}
\usepackage{natbib}
\usepackage{graphicx,subfigure}  
\usepackage{epstopdf}
\usepackage{dcolumn}  
\usepackage{bm}          
\usepackage{slashed}
\usepackage[utf8]{inputenc}
\usepackage{CJK}
\usepackage{mathptmx}
\usepackage{mathrsfs}
\usepackage[normalem]{ulem} 
\usepackage[dvipsnames]{xcolor} 
\usepackage{array}
\usepackage{multirow}

\begin{document}
\title{Investigating the muon catalyzed fusion in muonic lithium hydride}

\author{Qian Wu}
\affiliation{Institute of Modern Physics, Chinese Academy of Sciences, Lanzhou 730000, China}

\author{Xurong Chen}~\email{xchen@impcas.ac.cn}
\affiliation{Institute of Modern Physics, Chinese Academy of Sciences, Lanzhou 730000, China}
\affiliation{School of Nuclear Science and Technology, University of Chinese Academy of Sciences, Beijing 100049, China}
\affiliation{Guangdong Provincial Key Laboratory of Nuclear Science, Institute of Quantum Matter,
South China Normal University, Guangzhou 510006, China}

\begin{abstract}
In this work, we consider the muonic LiH as a new stage of the $\mu$ catalyzed fusion.
We calculate the bound states of the muonic LiH and LiH$^+$ and their wave functions based on the three and four body approximation.
In order to solve the Schr$\ddot{\rm o}$dinger equation, we apply the Gaussian expansion method which
gives us both the eigen energies and the wave functions.
The existences of the bound states in muonic LiH and LiH$^+$ are confirmed.
We find the dynamical behaviors between the $^7$Li and proton in muonic LiH and LiH$^+$ are similar as the ones in dt$\mu_2$ and dt$\mu$,
respectively.
The $\alpha$ sticking probability
in the muonic LiH$^+$ after the nuclear fusion is studied.
The calculated $\alpha$ sticking probability of being captured to the ground state of He$\mu$ is 5 times smaller than the one in dt$\mu$.

\end{abstract}
\maketitle
\section{Introduction}
The nuclear fusion, known as the nuclear reaction between the light nuclei, gives out a few MeVs energy due to the mass difference between the initial and final nuclei.
Thus, it is considered to be an important way to obtain nuclear power from nature.
However, in order to make the nuclear fusion happen within the nuclear distance (a few fms), the coulomb barrier
between two nuclei has to be overcome which requires a high temperature or pressure. And by far, no design of nuclear fusion has produced more fusion energy than the input electrical power.

The $\mu$ catalyzed fusion ($\mu$CF) has been studied as a reliable low energy nuclear fusion for a long history \cite{Jackson1957,Zelddovich1960,Vesman,kamimura1988,kamimura1993,Kami2022,Kami1989}. Among all of the $\mu$CF studies, the dt$\mu$ system is considered to be a possible resource of fusion which happens in a deuterium and
tritium (D-T) mixture.
The idea of $\mu$CF originates from the fact that the $\mu$ mass (105.66 MeV) is almost 207 times of the electron mass (0.55 MeV). Then, the radius of the $\mu^-$ atom is nearly 1/207 times smaller than the normal atom. Considering a molecular consisting of the deuteron, triton and $\mu$, the distance between two nuclei should be much smaller than a normal D$_2$ or T$_2$ molecular.

The first key process of dt$\mu$ $\mu$CF is the formation of the t$\mu$ atom. When the $\mu$s go into the D-T mixture, they will immediately ($\sim 10^{-11} \mathrm{~s}$) replace the electrons inside the deuterium and tritium to formate the d$\mu$ and t$\mu$ due to a much larger binding energy with the deuteron and triton. More importantly, since the mass of the triton is heavier than the deuteron, the binding energy of t$\mu$ is larger than the d$\mu$. Thus, the muon inside d$\mu$ will be taken by the triton to form t$\mu$, which is known as the so-called muon transfer reaction \cite{Hiyama2003GEM}.

Then, the t$\mu$ atom, which is electrically neutral, enters into a D$_2$ molecule and is captured by a deuteron to form a dt$\mu$ molecule.
It should be noted that the formation of the dt$\mu$ molecular happens via the Vesman mechanism \cite{Vesman} in ${\rm t} \mu(2 s)-{\rm D}_{2}$ scattering,
$\mathrm{t}\mu(2 \mathrm{~s})+\mathrm{D}_{2} \rightarrow[(\mathrm{dt} \mu) \mathrm{dee}]$.
If the energy of the dt$\mu$ molecular with respect to the t$\mu$-d threshold is less than the
dissociation energy of ${\rm D}_{2} \simeq 4.56$ eV, the released energy during the formation of the dt$\mu$ can be absorbed by the electrons in the D$_2$ which make it transfer into the excited states without breaking the whole D$_2$ molecular. In Refs. \cite{Vesman,dtmform1982}, they gave the muon molecular's formation rate of
$10^8$ s$^{-1}$ which is almost 200 times larger than the muon decay rate.

Finally, in the dt$\mu$ molecule, fusion reaction ${\rm d}+{\rm t} \rightarrow \alpha+{\rm n}+17.6$ MeV takes place immediately $\left(\sim 10^{-12}
\mathrm{~s}\right)$ due to the small distance between deuteron and triton \cite{Kami2022,Kami1989}. After that, the $\mu$ becomes free again and continues to catalyze the fusion reaction until its life time is exhausted.
Unfortunately, with a small probability $(\sim 1\%)$ \cite{alphastick1986,Zelddovich1960}, $\mu$ is captured by the generated $\alpha$ particle from the fusion reaction and exhausts its lifetime in the atom, although there is a probability of that the muon could be recaptured by the D$_2$ or T$_2$ during the time sticking with the $\alpha$ particle.
This sticking behavior of the $\mu$ causes a major reduction of the fusion times for each muon.

In order to produce one $\mu$, the input energy is estimated to be $\sim 5$ \rm{GeV} \cite{1980Petrov}. If $N_{f}$ is the number of fusions catalyzed by one muon, since one d-t fusion generates 17.6 MeV, we easily judge that $N_{f}\sim 280$ is necessary to reach the scientific break-even. Experiments performed so far show that $N_{f}$ increases almost linearly with the density of the $\mathrm{D}_{2} / \mathrm{T}_{2}$ mixture, reaching $N_{f} \sim 150$ at the density of liquid hydrogen. However, the fusion number comes to a limitation if we can't solve the alpha sticking problem.

In terms of the alpha sticking issue, we consider a muonic lithium hydride (LiH) molecular as the fusion reaction stage instead of the dt$\mu$.
When a muonic LiH or LiH$^+$ is formed, the fusion reaction $^7{\rm Li}+{\rm p}\rightarrow2\alpha+17.35$ MeV could happen.
According to Ref. \cite{Jackson1957}, the alpha sticking probability strongly depends on the velocity of the outgoing alpha particle.
In the nuclear fusion reaction $^7{\rm Li}+{\rm p}\rightarrow2\alpha$, the outgoing alpha particle carries more momentum than the one in the d+t$\rightarrow\alpha$+n. In this case, it may be more difficult for the alpha particle to capture the muon than in the D-T mixture.
Besides, since the $^7$Li$^{3+}$ has three plus electric charge and the muon may have a large probability to be recaptured by the $^7$Li$^{3+}$ with the muon transfer reaction after the fast moving He$-\mu$ is slowed down.

In this work, we investigate the alpha sticking probability after the fusion reaction happens in the muonic LiH and LiH$^+$.
First, we study the bound system of the Lip$\mu_4$ and Lip$\mu_3^+$.
In order to solve the few body Schr$\ddot{\rm o}$dinger equation, we apply the Gaussian expansion method (GEM), which is a reliable variational method and is used in nuclear and atomic physics \cite{Hiyama2003GEM,35hiyama1997ptp}.
The strict five body or six body Li+p+$\mu$s systems are difficult to solve with GEM, however, due to
the extremely large number of Gaussian basis which goes beyond our temporary supercomputer's power.
Then, we take an approximation.
As we know, the first ionization energy of the Li is 5.39 eV while the second and third
ionization energy are 75.64 eV and 122.45 eV, respectively.
Thus, we give a vague picture of the bound Lip$\mu_4$ and Lip$\mu_3$ molecular by approximately treat the Li$\mu_2$ as a whole part.
The bound system of muonic LiH and LiH$^+$ are discussed in the next section.
Second, in section \uppercase\expandafter{\romannumeral3}, we calculate the alpha sticking probability in muonic LiH$^+$
after the fusion reaction.
After that, we give a summary in the final section.

\section{Bound state of muonic LiH and LiH$^+$}
\begin{figure}[htbp]
\setlength{\abovecaptionskip}{0.cm}
\setlength{\belowcaptionskip}{-0.cm}
\centering
\includegraphics[width=0.45\textwidth]{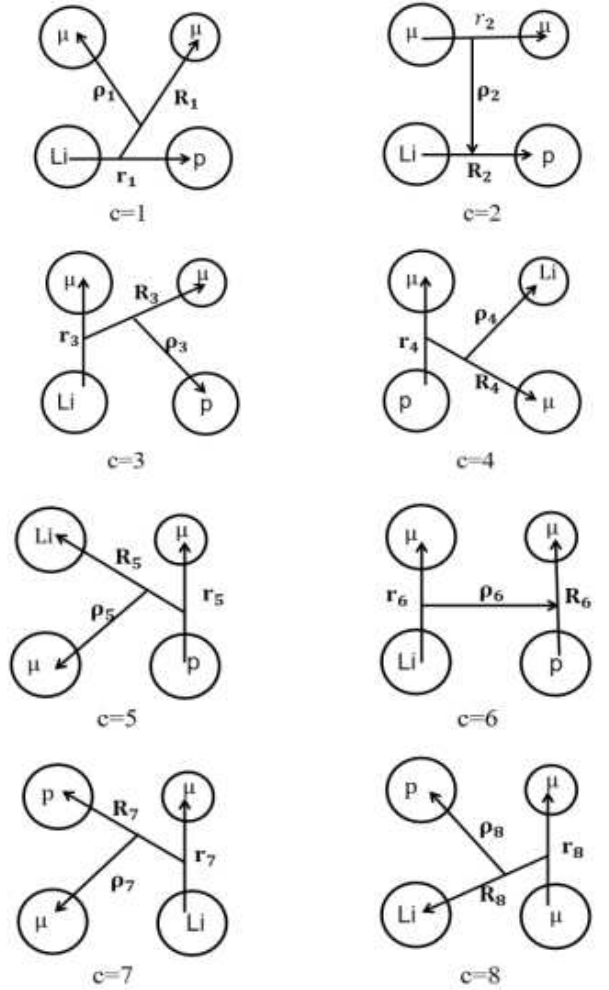}
\caption{
Jacobian coordinates of four body Li$\mu_2$-p-$\mu$-$\mu$ system. Here Li stands for the Li$\mu_2$ in short.
}
\label{fig:jaco}
\end{figure}

As we mentioned in the first section, we treat the muonic LiH as a four body Li$\mu_2$-p-$\mu$-$\mu$ system and the
muonic LiH$^+$ as a three body system. Then, the four body hamiltonian is written as:
\begin{equation}
H=T+\frac{e^{2}}{r_{1}}+\frac{e^{2}}{r_{2}}-\frac{e^{2}}{r_{3}}-\frac{e^{2}}{r_{5}},
\end{equation}
where $T$ is the kinetic energy and $r_1~r_5$ are the relative coordinates in Fig. \ref{fig:jaco}.
Then, in order to solve this equation, we apply the Gaussian expansion method and the trivial wave function is constructed with a
sum of 8 Jacobian coordinates channels, shown in Fig. \ref{fig:jaco} and the whole Li$\mu_2$ part is named as the Li in short.
The total wave function of the munic LiH with angular momentum J and its z-component is written as follows:
\begin{equation}\label{eq1}
\begin{aligned}
\Phi_{JM}({\rm LiH})&=\sum_{c=1}^{8} \sum_{ n \ell N L v \lambda \Lambda I S} C^c_{ n \ell N L v \lambda \Lambda I S}
\mathcal{A}[[[\psi_{n \ell}(\mathbf{r_c}) \\
 & \otimes \phi_{N L}(\mathbf{R_c})]_{\Lambda}\otimes \varphi_{\nu \lambda}(\rho_c)]_{I}
[\chi^{\mu_1}_{1/2}\chi^{\mu_2}_{1/2}]_{S}]_{JM}
\end{aligned}
\end{equation}
where the $C^c_{n \ell N L v \lambda \Lambda I S}$ are the parameters and $\mathcal{A}$ is the antisymmetric operator between the two muons.
$\chi^{\mu}_{1/2}$ is the spin wave function of the muon. In this work, we omit the spin of the all nuclei we mention.
The basis functions have the Gaussian radial shape multiplied by spherical harmonics.
\begin{equation}
\psi_{i \ell}(\mathbf{r})=r^{\ell} e^{-\alpha_{i} r^{2}} Y_{\ell m}(\hat{\mathbf{r}}),
\end{equation}
and similarly for $\phi(\mathbf{R})$ and $\chi(\rho)$.
The Gaussian range parameters are taken to lie in a geometrical progression.
The parameters C are then obtained with applying Eq. \ref{eq1} into the Rayleigh-Ritz variational method.
As for the three body Li$\mu_2$-p-$\mu$ system, its wave function can be written as:
\begin{equation}\label{eq1-1}
\begin{aligned}
\Phi_{JM}({\rm LiH^+})=\sum_{c=1,3} \sum_{ n \ell N L} C^c_{ n \ell N L}
[\psi_{n \ell}(\mathbf{r_c})\otimes\phi_{N L}(\mathbf{R_c})]_{JM}
\end{aligned}
\end{equation}
where c=1 and 3 are the similar ones in Fig. \ref{fig:jaco} with taking $\rho_1$ and $\rho_3$ out.
Here we omit the spin of the single $\mu$.

It should be noted that we calculate the binding energy of the muonic LiH with respect to the Li$\mu$-p$\mu$ threshold, which is $-2763.1$ eV for Li$\mu_2$-$\mu$ and $-2528.9$ eV for p-$\mu$. And the binding energy for muonic LiH$^+$ is calculated with respect to the Li$\mu_2$-$\mu$.
Here, the binding energy between the Li$\mu_2$ and $\mu$ shows one major short come of our approximation.
As shown in Table.\ref{tab:level1}, the whole four body calculation of Li$\mu_3$ gives the first ionization energy with 861.4 eV while it gives 2763.1 eV when we treat the Li$\mu_2$ as a whole.
It is reasonable since the electrons inside Li has a 1S$_2$2S$_1$ structure.

For comparison, we calculate the ionization energy of Lithium atom and muonic Lithium atom, which are shown in Table. \ref{tab:level1}.
The first, second and third ionization energy of Lithium atom are close to the experimental values. As for the muonic Lithium atom,
the 2763.1 eV, which is calculated within two body Li$^+$-$\mu$ system, certainly gives the energy in the 1S orbit, which is not far from the 4 times of the first ionization energy of Li$\mu_3$, 861.5 eV.
The way we treat the four body Li$\mu\mu\mu$ calculation is
given in the next section.
\begin{table}[tbh]
\begin{center}
\caption{The ionization energy of Lithium atom and muon Lithium atom. The experimental values of Lithium atom
are taken from Ref. \cite{ionization}. The units are in eV.}
\begin{tabular}{p{2.0cm}p{2.0cm}p{2.5cm}p{2.0cm}}
  \hline
    & Li($e_3$)  & Li($\mu_3$) & Exp. \cite{ionization}\\
 \hline
 First   & 5.36   &  861.50  &5.392    \\
 Second  & 75.64  & 15369.00 &75.640   \\
 Third   & 122.44 & 24916.50 &122.454  \\
 Total   & 203.44 & 41146.99 &203.486  \\
 \hline
\end{tabular}
\label{tab:level1}
\end{center}
\end{table}

Once we treat the Li$\mu_2$ as a whole part, the structures of muonic LiH and LiH$^+$ may be similar as the dt$\mu_2$ and dt$\mu$, respectively.
Since they are all composed of two heavy and positive charged particles and two (one) light and negative charged particles.
This also reminds of the D$_2$ and H$_2$ molecular where the two hydrogen atoms are bound with the covalent bond and the two electrons are shared by the two protons.

In Table. \ref{tab:level}, we show the binding energy of the Lip$\mu_4$, Lip$\mu_3$, dt$\mu_2$ and st$\mu$. And we only show the ground states' energy (J=0).
The calculated binding energy of muonic LiH is 478.5 eV with respect to the Li$\mu_3$-p$\mu$ threshold and the binding energy of muonic LiH$^+$ is 207.1 eV.
Together with this, the binding energies of the ground state of dt$\mu_2$ and dt$\mu$ are also shown and all three states have the same angular momentum as J=0.
The binding energy for the dt$\mu$ is 319.1 eV with respect to the t$\mu$-d threshold.
And the binding energy for the doubly $\mu$ dt$\mu_2$ molecular is
536.8 eV, with respect to the t$\mu$-d$\mu$ threshold. This also agrees with the one calculated in Ref. \cite{kamimura2001}.
\begin{table}[tbh]
\begin{center}
\caption{Binding energies of the ground state of the Lip$\mu_4$, Lip$\mu_3$, dt$|mu_2$ and dt$\mu$.
Here d, t and Li represent the deuteron ($^2$H), triton ($^3$H) and $^7$Li nucleus, respectively.
$\Psi(0)$ is the wave function with the relative distance between two nuclei being equal to 0.
The $a_\mu$ represents the $\mu$ atomic unit which is equal to $\hbar^2/e^2m_\mu$.}
\begin{tabular}{p{2.8cm}p{2.8cm}p{2.8cm}}
  \hline
 Molecular   & $E_{00}$ (eV)  & $\Psi(0)$ ($a_{\mu}^{-3/2}$) \\
 \hline
 Lip$\mu_4$   & 478.5   &   0.016       \\
 Lip$\mu_3$   & 207.1   &   0.003       \\
 dt$\mu_2$    & 536.8   &   0.021      \\
 dt$\mu$      & 319.1   &   0.003   \\
 \hline
\end{tabular}
\label{tab:level}
\end{center}
\end{table}

\begin{figure}[htbp]
\setlength{\abovecaptionskip}{0.cm}
\setlength{\belowcaptionskip}{-0.cm}
\centering
\includegraphics[width=0.45\textwidth]{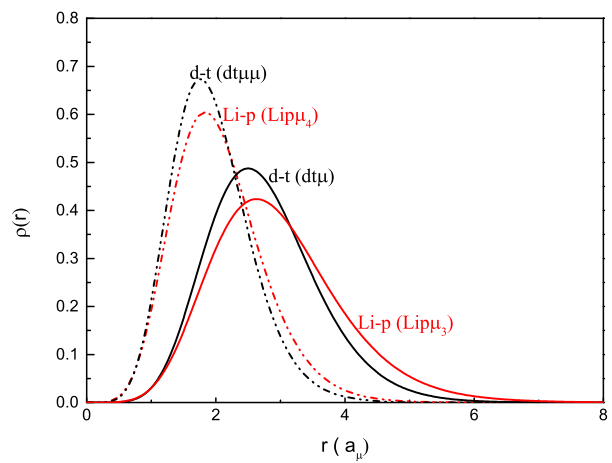}
\caption{
Density distributions $\rho(r)$ as a function of the relative distance between Li and proton (black) and between deuteron and triton (red).
The black solid and dash-dotted line represent the density distribution of the Lip$\mu_3$ and Lip$\mu_4$, respectively.
The red solid and dash-dotted line represent the density distribution of the dt$\mu\mu$ and dt$\mu$, respectively.
$a_\mu$ represents the $\mu$ atomic unit which is equal to $\hbar^2/e^2m_\mu$.}
\label{fig:density}
\end{figure}

It is reasonable to assume that the wave functions between the two nuclei in muonic LiH and dt$\mu_2$ are more compact than ones in the dt$\mu$
and muonic LiH$^+$. In the third column of table \ref{tab:level}, we show $\Psi(0)$ of these four systems which $\Psi(0)$
is defined as follows:
\begin{equation}
\Psi(0)=\int dR_1d\rho_1 \Psi(r_1=0,R_1,\rho_1).
\end{equation}
The $\Psi(0)$ in Lip$\mu_4$ and dt$\mu_2$ is larger than the ones in dt$\mu$ and muonic LiH$^+$.
Moreover, we calculate the density
distribution $\rho(r)$ of these four systems which is defined as:
\begin{equation}
\rho(r)=\int d\hat{r}dRd\rho \Psi(r,R,\rho).
\end{equation}
In Fig. \ref{fig:density}, the density distributions as a relative distance between Li$\mu_2$ and the proton in muonic LiH and LiH$^+$ are shown in the red solid and dash-dotted line. And the ones between deuteron and triton in dt$\mu_2$ and dt$\mu$ are shown in the black dashed and dash-dotted line. The density distribution and the $\Psi(0)$ of muonic LiH are similar as the ones in dt$\mu_2$. And similar behavior
exists between the muonic LiH$^+$ and dt$\mu$.

It is obvious that the major part of the wave function is located around 1 to 3 $a_\mu$ which is approximately around 100 to 500 fm. This allows a much more significant wave function around the 1 to 10 fm than the normal molecular.
In terms of the exact nuclear fusion rate, we plan to discuss it in our future work.

\section{$\alpha$ sticking probability}
In this section, we calculate the $\alpha$ sticking probability after the nuclear fusion reaction happens in the muonic LiH$^+$.
As we mentioned in the first section, our computer resource can't afford a full five body A$\mu\mu\mu\mu$ calculations.
In this case, we only calculate the sticking probability in the muonic LiH$^+$ and
solve a four body A$\mu\mu\mu$ system.

According to Ref. \cite{Zelddovich1960}, when the nuclear fusion happens, the nucleus $^7$Li and proton becomes close enough. Then,
the whole system can be regard as a core nucleus with the mass equal to Li+p and the positive charge 4.
At this moment, the wave function of
one single $\mu$ is given with solving the four body A$\mu_4$ system, where A$^{4+}$ is a combination of the $^7$Li and the proton.

When the fusion reaction:
\begin{equation}
^7\rm{Li}+\rm{p}\rightarrow2\alpha,\;  Q=17.35\;\rm{MeV}
\end{equation}
happens, the two $\alpha$ particles receive a recoil energy E and spread with a velocity v. Thus, when the $\mu$ is captured by the
fast-moving $\alpha$ particle to its ground state, its final state is
\begin{equation}
\phi_f(\bf{r})=e^{ip\cdot \bf{r}/\hbar}(\frac{8}{\pi a_\mu^3})^{1/2}e^{2r/a_\mu}.
\end{equation}
Here $p=m_\mu v$ and $r$ is the relative distance between the $\mu$ and the $\alpha$ particle.
Since the function $e^{ip\cdot \bf{r}/\hbar}$ oscillates strongly, the major contribution comes from the short-range part of $r$ ($r\leq\hbar/p$).
And the wave function at $r=0$ is propagational to $1/n^3$. One the other hand, the item $r^l$ in the excited states ($l\neq0$) should strongly reduce the integration than the $r^0$. Thus, the probability of being captured to the excited state of He-$\mu$ is much smaller than being captured to the ground state. Therefore, in this work, we only study the alpha sticking probability of $\mu$ being captured to the ground state of He$-\mu$.

The sticking probability S is given as:
\begin{equation}\label{eqa1}
S=\left|\int d\bf{r}^3\phi_f(\bf{r})\varphi_\mu(r)\right|^2
\end{equation}
where $\varphi_\mu(\bf{r})$ is the wave function of the ground state of $\mu$ particle in the A$\mu_4$.
In order to solve this equation, we write the oscillation part $e^{ip\cdot \bf{r}/\hbar}$ as:
\begin{equation}\label{eqa2}
e^{i\rm{p}\cdot \bf{r}/\hbar}=4\pi\sum_{\lambda=0}^{\infty}\sum_{\beta=-\lambda}^{\lambda}
(-i)^\lambda j_\lambda(-|r||p|/\hbar)Y^*_{\lambda\beta}(\hat{\bf{p}})Y_{\lambda\beta}(\hat{\bf{r}})
\end{equation}
where j$_\lambda(x)$ is the spherical Bessel function.
Thus, the integration over the $\bf{r}$ leads to a nonzero value only when $\lambda$=$\beta$=0.
Then, Eq. \ref{eqa1} turns to
\begin{equation}\label{eqa3}
S=4\pi\int_{0}^{\infty}\varphi_\mu(r)(\frac{8}{\pi a_\mu^3})^{1/2}e^{2r/a_\mu}\frac{sin(pr/\hbar)}{p\hbar}rdr.
\end{equation}
Besides, the relation between the p and the recoil energy E is
\begin{equation}\label{eqa4}
p=\sqrt{2EM_\mu^2/M_\alpha}.
\end{equation}
where M$_\mu$ and M$_\alpha$ are the mass of the $\mu$ and $\alpha$ particle, respectively.

\begin{figure}[htbp]
\setlength{\abovecaptionskip}{0.cm}
\setlength{\belowcaptionskip}{-0.cm}
\centering
\includegraphics[width=0.45\textwidth]{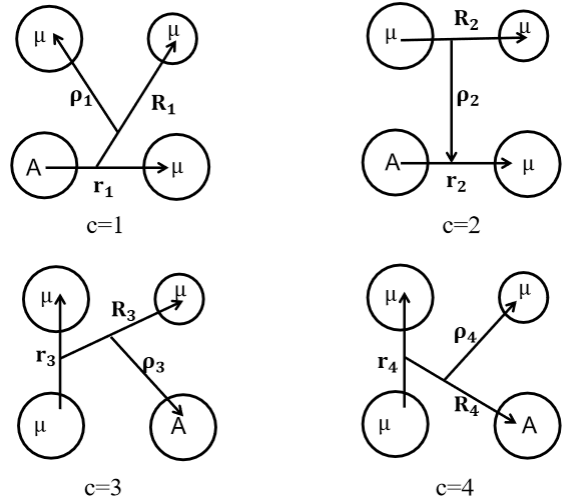}
\caption{
Jacobian coordinates of four body A-$\mu$-$\mu$-$\mu$ system.
}
\label{fig:jaco2}
\end{figure}

In order to obtain the wave function of the $\mu$, we need to solve the four body A$\mu_4$ system. Here we use the Gaussian
expansion method and the wave function is written as
\begin{equation}
\begin{aligned}
\Phi_{JM}&=\sum_{c=1}^{4} \sum_{n \ell N L v \lambda \Lambda I s S} C^c_{n \ell N L v \lambda \Lambda I s S}
\mathcal{A}[[[\psi_{n \ell}(\mathbf{r_c}) \\
 & \otimes \phi_{N L}(\mathbf{R_c})]_{\Lambda}\otimes \chi_{\nu \lambda}(\rho_c)]_{I}
 [[\chi^{\mu_1}_{1/2}\chi^{\mu_2}_{1/2}]_{s}\chi^{\mu_3}_{1/2}]_S]_{JM}.
\end{aligned}
\end{equation}
where c$=1\sim4$ is the channels of the Jacobian coordinates in Fig.\ref{fig:jaco2} and $\mathcal{A}$ is the antisymmetric operator between the three $\mu$s.
In Fig. \ref{fig:jaco2}, the A represents the heavy nuclei which is $^7$Li$+$p.
Accordingly, the wave function of one single $\mu$ is defined as
\begin{equation}
\varphi_\mu(r)=\left|\int d\hat{r_1}dR_1d\rho_1 \Phi(r,R,\rho)\right|^{1/2}
\end{equation}
And as mentioned in the former section,
the same mode space and method are used in
calculating the ionization energies of Lithium atom and muonic Lithium atom with simply replacing A with Li$^{3+}$.

We then calculate the alpha stick probability of the following two nuclear reactions:
\begin{equation}
\begin{aligned}
^7\rm{Li}+\rm{p}\rightarrow& 2^4\rm{He},\quad   E=8.67\;\rm{MeV}.\\
d+t \rightarrow& ^4\rm{He}+n,\quad   E=3.5\;\rm{MeV}.
\end{aligned}
\end{equation}
where E is the recoil energy of the $\alpha$ particle.
It should be noted that in the dt$\mu$, the wave function of the $\mu$ when the dt fusion happens can be regarded as the wave function of
the Helium atom \cite{Zelddovich1960}.

Shown in Table. \ref{tab:stick}, the $\alpha$ sticking probabilities of these two reactions are calculated to be 0.0098 and 0.0021, respectively for dt$\mu$ and Lip$\mu_3$.
As we can see, the sticking probability is almost five times smaller than the one in the dt$\mu$. As we mentioned in the introduction section,
the sticking probability have a strong dependence of fourth power of the velocity of the outgoing $\alpha$ particle \cite{Zelddovich1960}.
Thus, it is reasonable to have a much smaller sticking probability in LiH since the recoil energy in the fusion reaction is much larger than the one in dt$\mu$.

\begin{table}[tbh]
\begin{center}
\caption{Alpha sticking probability of dt$\mu$ and LiH$\mu_4$.}
\begin{tabular}{p{3cm}p{3cm}}
  \hline
 system   & probability    \\
 \hline
  dt$\mu$  & 0.0098        \\
 Lip$\mu_3$    & 0.0021    \\
 \hline
\end{tabular}
\label{tab:stick}
\end{center}
\end{table}

\section{Summary}
In order to give a possible solution to the $\alpha$ sticking issue in the normal dt$\mu$ $\mu$CF,
we put forward the muonic LiH and LiH$^+$ as the stage of the $\mu$CF.
The goal fusion reaction is $^7{\rm Li}+{\rm p}\rightarrow2\alpha+17.35$ MeV.
We calculate the ground state energies and the wave functions of the muonic LiH and LiH$^+$ based on a four body
Li$\mu_2+{\rm p}+\mu+\mu$ and three body Li$\mu_2+{\rm p}+\mu$ model.
To make the calculation practicable, we treat the Li$\mu_2$ as one whole part due to the large binding
energy of the two $\mu$s and the Li$^{3+}$.
We find that the dynamical behaviors between $^7$Li and proton in muonic LiH and LiH$^+$ are similar as the ones in dt$\mu_2$ and dt$\mu$, respectively.
Thus, the nuclear reaction in the muonic LiH and LiH$^+$ might happen immediately when the muonic LiH or LiH$^+$ is formed.
It should be noted that another nuclear reaction $^7{\rm Li}+{\rm p}\rightarrow ^7{\rm Be}+n$ could happen in muonic LiH.
And the detailed investigation concerning the fusion channels and rates in muonic LiH will be our next work.

Then, we calculate the $\alpha$ sticking probability after the fusion reaction.
We find that the sticking probability of the $\mu$ being captured to the ground state of He$\mu$ in the muonic LiH$^+$ is almost five times smaller than the one in the dt$\mu$.
However, it may not ease the current circumstance of the $\mu$CF concerning the $\alpha$ sticking issue as we explained in the first section.
We give further discussions as follows.
On the one hand, in the muonic LiH or LiH$^+$, it requires four or three $\mu$s
and two $\alpha$ particles are generated after the fusion.
Therefore, the total improvement of to the whole $\mu$CF cycle in muonic LiH or LiH$^+$ could be marginal.
On the other hand, if the He$-\mu$ is slowed down in
the LiH system, the $\mu$ particle could be recaptured by the Li$^{3+}$ with the $\mu$ transfer reaction.
The similar reaction also happens in d$\mu$-t$\mu$ transfer reaction \cite{Hiyama2003GEM}.
This calculation of the $\mu$ transfer reaction between the He$\mu$ and Li$\mu$ is our next work.

We have to admit that one major impediment of our conjecture is that the formation of the muonic LiH or LiH$^+$.
Since LiH$^{4+}$ has four positive charge, it has to capture at least three $\mu$s which we think is very difficult in the actual design.
But we hope modern experimental technology can help solve this problem. 

\begin{acknowledgments}
This work is supported by the Strategic Priority Research Program of Chinese Academy of Sciences under the Grant NO. XDB34030301, the National Natural Science Foundation of China under the Grant NOs. 12005266, 12075288, 11735003, 11961141012 and Guangdong Major Project of
Basic and Applied Basic Research No. 2020B0301030008. It is also
supported by the Youth Innovation Promotion Association CAS.
\end{acknowledgments}

\end{document}